\documentclass[sigconf]{acmart}

\usepackage{booktabs}
\usepackage{graphicx}
\usepackage{times}
\usepackage{helvet}
\usepackage{courier}
\usepackage{booktabs}
\usepackage{natbib}
\usepackage{epsfig}
\usepackage{amssymb}
\usepackage{amsmath}
\usepackage{amsfonts}
\usepackage{breqn}
\usepackage{multirow}
\usepackage{array}
\usepackage{subcaption}
\usepackage{mathtools}
\usepackage{caption}
\usepackage{url}
\usepackage{float}
\usepackage{balance}
\usepackage{algorithm}
\usepackage{algorithmic}
\usepackage{pifont}
\usepackage{flushend}

\newcommand{\bul}[1]{\underline{{#1}}}
\newcommand{\eq}[1]{Eq.~(\ref{#1})}
\newcommand{\xhdr}[1]{\vspace{0.4mm}\noindent{{\bf #1}}}
\newcommand{\md}{\emph{TransRec}}
\newcommand{\modelnamelong}{\emph{Translation-based Recommendation}}
\DeclareMathOperator*{\argmax}{arg\,max}

\begin{document}

\copyrightyear{2017} 
\acmYear{2017} 
\setcopyright{acmcopyright}
\acmConference{RecSys '17}{August 27-31, 2017}{Como, Italy}\acmPrice{15.00}\acmDOI{10.1145/3109859.3109882}
\acmISBN{978-1-4503-4652-8/17/08}

\title{Translation-based Recommendation}

\author{Ruining He}
\affiliation{%
  \institution{UC San Diego}
}
\email{r4he@cs.ucsd.edu}

\author{Wang-Cheng Kang}
\affiliation{%
  \institution{UC San Diego}
}
\email{wckang@eng.ucsd.edu}

\author{Julian McAuley}
\affiliation{%
  \institution{UC San Diego}
}
\email{jmcauley@cs.ucsd.edu}

\begin{abstract}
Modeling the complex interactions between users and items as well as amongst items themselves is at the core of designing successful recommender systems. One classical setting is predicting users' personalized sequential behavior (or `next-item' recommendation), where the challenges mainly lie in modeling `third-order' interactions between a user, her previously visited item(s), and the next item to consume. 
Existing methods typically decompose these higher-order interactions into a combination
of
\emph{pairwise} relationships, by way of which user preferences (user-item interactions) and sequential patterns (item-item interactions) are captured by separate components.
In this paper, we propose a unified method, \md{}, to model such third-order relationships for large-scale sequential prediction. Methodologically, we embed items into a `transition space' where users are modeled as \emph{translation} vectors operating on item sequences. Empirically, this approach outperforms the state-of-the-art on a wide spectrum of real-world datasets. 
Data and code are available at \url{https://sites.google.com/a/eng.ucsd.edu/ruining-he/}.
\end{abstract}

\maketitle
\section{Introduction}
Modeling and predicting the \emph{interactions} between users and items, as well as the \emph{relationships} amongst the items themselves are the main tasks of recommender systems. 
For instance, in order to predict \emph{sequential} user actions like the next product to purchase, movie to watch, or place to visit, it is essential (and challenging!) to model the \emph{third-order} interactions between a user ($u$), the item(s) she recently consumed ($i$), and the item to visit next ($j$). Not only does the model need to handle the 
complexity
of the interactions themselves, but also the 
scale and inherent sparsity of real-world data.

\begin{figure}[!t]
\centering
\includegraphics[width=\linewidth]{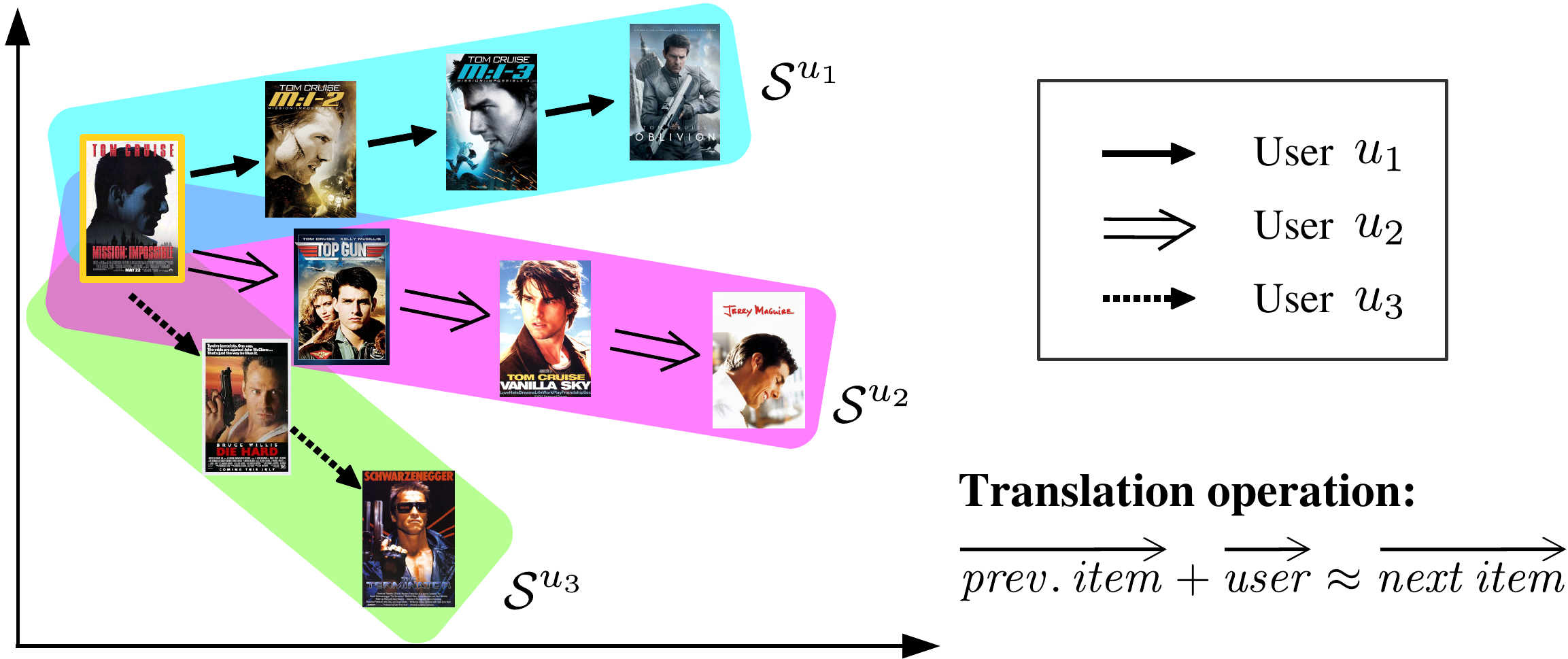}
\caption{\md{} as a sequential model: Items 
(movies)
are embedded into a `transition space' where each user is modeled by a \emph{translation} vector. The transition of a user from one item to another is captured by a user-specific translation operation. Here we demonstrate the historical sequences $\mathcal{S}^{u_1}$, $\mathcal{S}^{u_2}$, and $\mathcal{S}^{u_3}$ of three users. Given the same starting point, the movie \emph{Mission: Impossible I}, $u_1$ went on to watch the whole series, $u_2$ continued to watch drama movies by Tom Cruise, and $u_3$ switched to similar action movies.}
\label{fig:idea}
\end{figure}

Traditional recommendation methods usually excel at modeling two-way (i.e.,~pairwise) interactions. There are Matrix Factorization (MF) techniques \cite{koren2009matrix} that make use of inner products to model the compatibility between user-item pairs (i.e.,~user preferences). Likewise, there are also (first-order) Markov Chain (MC) models \cite{stochasticprocesses} that capture transition relationships between pairs of adjacent items in sequences (i.e.,~sequential dynamics), often by way of factorizing the transition matrix in favor of generalization ability. For the task of sequential recommendation, researchers have made use of scalable tensor factorization methods, such as Factorized Personalized Markov Chains (FPMC) proposed by Rendle \emph{et al.} \cite{rendle2010fpmc}. FPMC models third-order relationships between $u$, $i$, and $j$ by the \emph{summation} of two pairwise relationships: one for the compatibility between $u$ and the next item $j$, and another for the sequential continuity between the previous item $i$ and the next item $j$. Ultimately, this is a combination of MF and MC (see Section \ref{sec:connect} for details).

Recently, there have been two lines of works that aim to improve FPMC. Personalized metric embedding methods replace the inner products in FPMC with Euclidean distances, where the metricity assumption---especially the triangle inequality---enables the model to generalize better \cite{wu2013personalized,moore2013taste,feng2015prme}. However, these works still adopt the framework of modeling the user preference component and sequential continuity component separately, which may 
be disadvantageous as the two components are inherently correlated.

Another line of work \cite{hrm} makes use of operations like average/max pooling to \emph{aggregate} the representations of the user $u$ and the previous item $i$, before their compatibility with the next item $j$ is measured. These works partially address the issue of modeling the dependence of the two key components, though are hard to interpret and can not benefit from the generalization ability of metric embeddings.

In this paper, we aim to tackle the above issues by introducing a new framework called \modelnamelong{} (\md). The key idea behind \md{} is presented in Figure \ref{fig:idea}: Items are embedded as points in a (latent) `transition space'; each user is represented as a `translation vector' in the same space. Then, the third-order interactions mentioned earlier are captured by a personalized translation operation: the coordinates of previous item $i$, plus the translation vector of $u$ determine (approximately) the coordinates of the next item $j$, i.e.,~$\vec{\gamma}_i + \vec{t}_u \approx \vec{\gamma}_j$. Finally, we model the compatibility of the $(u,i,j)$ triplet with a distance function $d(\vec{\gamma}_i + \vec{t}_u, \vec{\gamma}_j)$. At prediction time, recommendations can be
via nearest-neighbor search centered at $\vec{\gamma}_i + \vec{t}_u$.

The advantages of such an approach are three-fold: (1) \md{} naturally models third-order interactions with only a \emph{single} component; (2) \md{} also enjoys the generalization benefits from the implicit metricity assumption; and (3) \md{} can easily handle large sequences (e.g.,~millions of instances) due to its simple form.
Empirically, we conduct comprehensive experiments on a wide range of large, real-world datasets (which are publicly available), and quantitatively demonstrate the superior recommendation performance achieved by \md.

In addition to the sequential prediction task, we also investigate the strength of \md{} at tackling item-to-item recommendation where pairwise relations between items need to be captured, e.g.,~suggesting a shirt to match a previously purchased pair of pants. State-of-the-art works for this task are mainly based on metric or non-metric embeddings (e.g.,~\cite{he2016monomer,VisualSIGIR}). We empirically evaluate \md{} on eight large co-purchase datasets from \emph{Amazon} and find it to significantly outperform multiple state-of-the-art models by using the translation structure.

Finally, we introduce a new large, sequential prediction dataset, from \emph{Google Local}, that contains a large corpus of ratings and reviews on millions of businesses around the world.

\section{Related Work}
\xhdr{General recommendation.} Traditional approaches to recommendation ignore sequential signals in the system. Such systems focus on modeling user preferences, and typically rely on Collaborative Filtering (CF) techniques, especially Matrix Factorization (MF) \cite{Handbook}. For implicit feedback data (like purchases, clicks, and thumbs-up), point-wise and pairwise methods based on MF have been proposed. Point-wise methods (e.g.,~\cite{WRMF,OCCF,ning2011slim}) assume all non-observed feedback to be negative and factorize the user-item feedback matrix. In contrast, pairwise methods (e.g.,~\cite{rendle2009bpr,rendle2010pairwise,rendle2014improving}) make a weaker assumption that users simply prefer observed feedback over unobserved feedback and optimize the pairwise rankings of (positive, non-positive) pairs. 

\xhdr{Modeling temporal dynamics.} Several works extend general recommendation models to make use of timestamps associated with feedback. For example, early similarity-based CF (e.g.,~\cite{timeweightCF}) uses time weighting schemes that assign decaying weights to previously-rated items when computing similarities. 
More recent efforts are mostly based on MF, where the goal is to model and understand the historical \emph{evolution} of users and items, e.g.,~Koren \emph{et al.} achieved state-of-the-art rating prediction results on \emph{Netflix} data, largely by exploiting temporal signals \cite{koren2010temporal,koren2009matrix}. The sequential prediction task we are tackling is related to the above, except that instead of directly using those timestamps, it focuses on learning the sequential relationships between user actions (i.e.,~it focuses on the \emph{order} of actions rather than the specific time).

\xhdr{Sequential recommendation.} Scalable sequential models usually rely on Markov Chains (MC) to capture sequential patterns (e.g.,~\cite{rendle2010fpmc,hrm,feng2015prme}). Rendle \emph{et al.} proposed to factorize the 
third-order `cube' that represents the transitions amongst items made by users. The resulting model, Factorized Personalized Markov Chains (FPMC), can be seen as a combination of MF and MC and achieves good performance for next-basket recommendation. 

There are also works that have adopted metric embeddings for the recommendation task, 
leading to better generalization ability.
For example, Chen \emph{et al.} introduced Logistic Metric Embeddings (LME) for music playlist generation \cite{chen2012playlist}, where the Markov transitions among different songs are encoded by the distances among them. Recently, Feng \emph{et al.} further extended LME to model personalized sequential behavior and used pairwise ranking for predicting next points-of-interest \cite{feng2015prme}. 
On the other hand, Wang \emph{et al.} recently introduced the Hierarchical Representation Model (HRM), which extends FPMC by applying aggregation operations (like max/average pooling) to model more complex interactions. We will give more details of these works in Section \ref{sec:seqmodels}.

Our work differs from the above in that we introduce a \emph{translation}-based structure which naturally models the third-order interactions between a user, the previous item, and the next item for personalized Markov transitions.

\begin{table}
\centering
\renewcommand{\tabcolsep}{5pt}
\caption{Notation} \label{tb:notation}
\begin{tabular}{ll} \toprule
Notation & Explanation\\ \midrule
$\mathcal{U}$, $\mathcal{I}$ & user set, item set \\
$u$, $i$, $j$ & user $u \in \mathcal{U}$, items $i, j \in \mathcal{I}$\\
$\mathcal{S}^u$ & historical sequence of user $u$: $(\mathcal{S}^u_1, \mathcal{S}^u_2, \cdots, \mathcal{S}^u_{|\mathcal{S}^u|})$ \\
$\Phi$ & transition space; $\Phi = \mathbb{R}^{K}$ \\
$\Psi$ & a subspace in $\Phi$; $\Psi \subseteq \Phi$ \\
$\vec{\gamma}_i$ & embedding vector associated with item $i$; $\vec{\gamma}_i \in \Psi$\\
$\vec{t}$ & (global) translation vector $\vec{t} \in \Phi$\\
$\vec{t}_u$ & translation vector associated with user $u$; $\vec{t}_u \in \Phi$\\
$\vec{T}_u$ & $\vec{T}_u = \vec{t} + \vec{t}_u$; $\vec{T}_u \in \Phi$\\
$\beta_i$ & bias term associated with item $i$; $\beta_i \in \mathbb{R}$ \\
$\vec{f}_i$ & explicit feature vectors associated with item $i$\\
$d(x, y)$ & distance between $x$ and $y$ \\
\bottomrule
\end{tabular}
\end{table}

\xhdr{Knowledge bases.} Although different from recommendation, there has been a large body of work in knowledge bases that focuses on modeling multiple, complex relationships between various entities. Recently, partially motivated by the findings made by word2vec \cite{mikolov2013word2vec}, translation-based methods (e.g.,~\cite{TransE,TransR,TransH}) have achieved state-of-the-art accuracy and scalability, in contrast to those achieved by traditional embedding methods relying on tensor decomposition or collective matrix factorization (e.g.,~\cite{nickel2011RESCAL,nickel2012YAGO,singh2008relational}). Our work is inspired by those findings, and we tackle the challenges from modeling large-scale, personalized, and complicated sequential data. This is the first work that explores this direction to the best of our knowledge.

\section{The Translation-based Model}
\subsection{Problem Formulation}
We refer to the objects that users ($\mathcal{U}$) interact with in the system as items ($\mathcal{I}$), e.g.,~products, movies, or places. The \emph{sequential}, or `next-item,' prediction task we are tackling is formulated as follows. For each user $u \in \mathcal{U}$ we have a sequence of items $\mathcal{S}^u = (\mathcal{S}^u_1, \mathcal{S}^u_2, \cdots, \mathcal{S}^u_{|\mathcal{S}^u|})$ that $u$ has interacted with. Given the sequence set from all users $\mathcal{S} = \{\mathcal{S}^{u_1}, \mathcal{S}^{u_2}, \cdots, \mathcal{S}^{u_{|\mathcal{U}|}}\}$, 
our objective is to predict the next item to be `consumed' by each user and generate recommendation lists accordingly. Notation used throughout the paper is summarized in Table \ref{tb:notation}.

\subsection{The Proposed Model} \label{sec:model}
We aim to build a model that (1) naturally captures personalized sequential behavior, and (2) easily scales to large, real-world datasets. 
Methodologically, we learn a transition space 
$\Phi = \mathbb{R}^{K}$,
where each item $i$ is represented with a point/vector $\vec{\gamma}_i \in \Phi$. $\vec{\gamma}_i$ can be \emph{latent}, or transformed from certain explicit features of item $i$, e.g.,~the output of a neural network. In this paper we take $\vec{\gamma}_i$ as latent vectors.

Recall that the historical sequence $\mathcal{S}^u$ of user $u$ is a series of transitions $u$ has made from one item to another. 
To model the \emph{personalized} sequential behavior, 
we represent each user $u$ with a \emph{translation} vector $\vec{t_u} \in \Phi$ to capture $u$'s inherent intent or `long-term preferences' that influenced her to make these decisions. In particular, if $u$ transitioned from item $i$ to item $j$, what we want is 
$$
\vec{\gamma}_i + \vec{t}_u \approx \vec{\gamma}_j,
$$
which means $\vec{\gamma}_j$ should be a nearest neighbor of $\vec{\gamma}_i + \vec{t}_u$ in $\Phi$ according to some distance metric $d(x,y)$, e.g.,~$\mathcal{L}_1$ distance.

Note that we are uncovering a metric space where (1) \emph{neighborhood} captures the notion of similarity and (2) \emph{translation} encapsulates various semantically complex transition relationships amongst items. In both cases, the inherent triangle inequality assumption plays an important role in helping the model to generalize well, as it does in canonical metric learning scenarios. 
For instance, if users tend to transition from item \emph{A} to two items \emph{B} and \emph{C}, then \md{} will also put \emph{B} close to \emph{C}.
This is a desirable property especially when data sparsity is a major concern. One plausible alternative is to use the inner product of $\vec{\gamma}_i + \vec{t}_u$ and $\vec{\gamma}_j$ to model their `compatibility.' 
However, this way item \emph{B} and \emph{C} in our above example might be far from each other because inner products do not guarantee the triangle inequality condition. 

Due to the sparsity of real-world datasets, it might not be affordable to learn separate translation vectors $\vec{t}_u$ for each user. 
Therefore we add another translation vector $\vec{t}$ to capture `global' transition dynamics across all users, and we let  
$$
\vec{T}_u = \vec{t} + \vec{t}_u.
$$
This way $\vec{t}_u$ can be seen as an offset vector associated with user $u$. 
Although doing so yields no additional expressive power,\footnote{Note that we can still learn personalized sequential behavior as users are being parameterized separately.} the advantage is that $\vec{t}_u$'s of \emph{cold-start} users will be regularized towards 0 and we are essentially using $\vec{t}$---the `average' behavior---to make predictions for these users. 

Finally, the probability that a given user $u$ transitions from the previous item $i$ to the next item $j$ is predicted by
\begin{equation} \label{eq:TransRec}
\begin{aligned}
&\mathit{Prob}(j~|~u, i) \propto  \beta_j -  d(\vec{\gamma}_i + \vec{T}_u, \vec{\gamma}_j), \\
&\text{subject to} \quad \vec\gamma_i \in \Psi \subseteq \Phi, \;\;\; \text{for}{\ } i \in \mathcal{I}.
\end{aligned}
\end{equation}
$\Psi$ is a subspace in $\Phi$, e.g.,~a unit ball, a technique which
has been shown to be helpful for mitigating `curse of dimensionality' issues (e.g.,~\cite{TransE,TransH,TransR}). 
In the above equation a single bias term $\beta_j$ is added to capture overall item popularity. 

\xhdr{Ranking Optimization.}
Given a user and the associated historical sequence, the ultimate goal of the task is to rank the ground-truth item $j$ higher than all other items ($j'\in \mathcal{I} \setminus j$). Therefore it is a natural choice to optimize the pairwise ranking between $j$ and $j'$ by (e.g.) Sequential Bayesian Personalized Ranking (S-BPR) \cite{rendle2010fpmc}. 
To this end, we optimize the total order $>_{u,i}$ given the user $u$ and the previous item $i$ in the sequence:
\begin{equation} \label{eq:obj}
\begin{split}
\widehat{\Theta} &= \argmax_{\Theta} ~ \ln \prod_{u \in \mathcal{U}} \prod_{j \in \mathcal{S}^u} \prod_{j'\notin \mathcal{S}^u} \mathit{Prob}(j >_{u,i} j' | \Theta) ~ \mathit{Prob}(\Theta) \\
&= \argmax_{\Theta} ~ \sum_{u \in \mathcal{U}} \sum_{j \in \mathcal{S}^u} \sum_{j'\notin \mathcal{S}^u} \ln \sigma(\widehat{p}_{u,i,j} - \widehat{p}_{u,i,j'}) - \Omega(\Theta),
\end{split}
\end{equation}
where $i$ is the preceding item\footnote{Here $j$ can not be the first item in the sequence $\mathcal{S}^u$ as it has no preceding item.} of $j$ in $\mathcal{S}^u$, $\widehat{p}_{u,i,j}$ is a shorthand for the prediction in \eq{eq:TransRec}, 
$\Theta$ is the parameter set $\{ \beta_{i\in \mathcal{I}}, \vec{\gamma}_{i\in \mathcal{I}}, \vec{t}_{u\in\mathcal{U}}, \vec{t}\}$, and
$\Omega(\Theta)$ is an $\mathcal{L}_2$ regularizer.
Note that according to S-BPR, the probability that the ground-truth item $j$ is ranked higher than a `negative' item $j'$ (i.e.,~$\mathit{Prob}(j >_{u,i} j' | \Theta)$) is estimated by the sigmoid function $\sigma(\widehat{p}_{u,i,j} - \widehat{p}_{u,i,j'})$.

\subsection{Inferring the Parameters}
\xhdr{Initialization.} Item embeddings $\vec{\gamma}_{i\in\mathcal{I}}$ and $\vec{t}$ are randomly initialized to be unit vectors. $\beta_{i\in\mathcal{I}}$ and $\vec{t}_{u\in\mathcal{U}}$ are initialized to be zero.

\xhdr{Learning Procedure.}
The objective function (\eq{eq:obj}) is maximized by stochastic gradient ascent: First, we uniformly sample a user $u$ from $\mathcal{U}$. Then, a `positive' item $j$ and a `negative' item $j'$ are uniformly sampled from $\mathcal{S}^u\setminus\mathcal{S}^u_1$ and $\mathcal{I}\setminus \mathcal{S}^u$ respectively. Next, parameters are updated via stochastic gradient ascent: 
$$
\Theta \leftarrow \Theta + \epsilon \cdot \left(\sigma(\widehat{p}_{u,i,j'}-\widehat{p}_{u,i,j}) ~ \frac{\partial (\widehat{p}_{u,i,j} - \widehat{p}_{u,i,j'})}{\partial \Theta} - \lambda_{\Theta} \cdot \Theta \right), 
$$
where $\epsilon$ is the learning rate and $\lambda_{\Theta}$ is a regularization hyperparameter. Finally, we re-normalize $\vec{\gamma}_i$, $\vec{\gamma}_j$, and $\vec{\gamma}_{j'}$ to be vectors in $\Psi$. For example, if we let $\Psi$ be the unit $\mathcal{L}_2$-ball, then $\vec{\gamma} \leftarrow \vec{\gamma} / \text{max}(1, \|\vec{\gamma}\|)$.
The above steps are repeated until convergence or until the accuracy plateaus on the validation set. 

\subsection{Nearest Neighbor Search} 
At test time, recommendation can be made via nearest neighbor search. A small challenge lies in handling bias terms: First, we replace $\beta_j$ with $\beta'_j = \beta_j - \max_{k \in \mathcal{I}} \beta_k$ for $j \in \mathcal{I}$. Shifting the bias terms does not change the ranking of items for any query. Next, we absorb $\beta'_j$ into $\vec{\gamma}_{j}$ and get $\vec{\gamma}'_{j} = (\vec{\gamma}_{j}; \sqrt{-\beta'_j})$ for (squared) $\mathcal{L}_2$ distance, or $\vec{\gamma}'_{j} = ( \vec{\gamma}_{j}; \beta'_j)$ for $\mathcal{L}_1$ distance. 
Finally, given a user $u$ and an item $i$, we obtain the `query' coordinate $(\vec{\gamma}_i + \vec{T}_u; 0)$, which can then be used for retrieving nearest neighbors in the space of $\vec{\gamma}'_{j}$.

\subsection{Connections to Existing Models} \label{sec:connect}
\subsubsection{Knowledge Graphs}
 Our method is inspired by recent advances in knowledge graph completion, e.g.,~\citep{TransE,TransH,TransR,DistMult,ComplexEmbedding}, where the objective is to model multiple types of {relations} between pairs of entities, e.g.,~Turing was born in England (`was\_born\_in' is the relation between `Turing' and `England'). One state-of-the-art technique (e.g.,~\citep{TransE,TransH,TransR}) is embedding entities as points and relations as \emph{translation} vectors such that the relationship between two entities is captured by the corresponding translation operation.
In the previous example, if we represent `Turing,' `England,' and `was\_born\_in' with vectors
$\overrightarrow{\mathit{head}}$, $\overrightarrow{\mathit{tail}}$, and $\overrightarrow{\mathit{relation}}$ respectively, then the following is desired: $\overrightarrow{\mathit{head}} + \overrightarrow{\mathit{relation}} \approx \overrightarrow{\mathit{tail}}$.

In recommendation settings, items are analogous to `entities' in knowledge graphs. Our key idea is to represent each user as one particular type of `relation' such that it captures the personalized reasons a user transitions from one item to another.

\subsubsection{Sequential Models} \label{sec:seqmodels}
State-of-the-art sequential prediction models are typically based on (personalized) Markov Chains. 
FPMC is a seminal model proposed by \cite{rendle2010fpmc}, whose predictor consists of two key components: (1) the inner product of user and item factors (capturing users' inherent preferences), and (2) the inner product of the factors of the previous and next item (capturing sequential dynamics). FPMC is essentially the combination of MF and factorized MC:
\begin{equation} \label{eq:fpmc}
\mathit{Prob}(j~|~u, i) \propto  \langle \vec{M}_u, \vec{N}_j \rangle + \langle \vec{P}_i, \vec{Q}_j \rangle,
\end{equation}
where user embeddings $\vec{M}_u$ and item embeddings $\vec{N}_j$, $\vec{P}_i$, $\vec{Q}_j$ are parameters learned from the data. 

Recently, Personalized Ranking Metric Embedding (PRME) \cite{feng2015prme} was proposed to improve FPMC by learning two metric spaces: one for measuring user-item affinity and another for sequential continuity. It predicts according to:
\begin{equation} \label{eq:prme}
\mathit{Prob}(j~|~u, i) \propto - \; \left(\alpha \cdot \| \vec{M}_u - \vec{N}_j \|_2^2 ~+~ (1 - \alpha) \cdot \| \vec{P}_i - \vec{P}_j \|_2^2 \right),
\end{equation}
which replaces inner products in FPMC by distances. As argued in \cite{chen2012playlist,feng2015prme}, the underlying metricity assumption brings better generalization ability. However, like FPMC, PRME still has to learn two closely \emph{correlated} components in a separate manner, 
using a hyperparameter $\alpha$ to balance them. 

Another recent work, Hierarchical Representation Model (HRM) \cite{hrm}, tries to extend FPMC by using an \emph{aggregation} operation (max/average pooling) to blend users' preferences ($\vec{M}_u$) and their recent activities ($\vec{N}_i$):
\begin{equation} \label{eq:hrm}
\mathit{Prob}(j~|~u, i) \propto  \langle \text{aggregation}(\vec{M}_u, \vec{N}_i), \vec{N}_j \rangle.
\end{equation}
Although the predictor can be seen as modeling the third-order interactions with a single component, the aggregation is hard to interpret and does not reap the benefits of using metric embeddings as PRME does. 
 
\md{} also falls into the category of Markov Chain models; however, it applies a novel \emph{translation}-based structure in a metric space, which enjoys the benefits of using a single, interpretable component as well as a metric space.

\section{Experiments}

\subsection{Datasets and Statistics}
To fully evaluate the capability and applicability of \md, in our experiments we include a wide range of publicly available datasets varying significantly in domain, size, data sparsity, and variability/complexity.

\begin{table}
\begin{center}
\setlength{\tabcolsep}{1.8pt}
\caption{Statistics (in ascending order of item density). }\label{tb:data}
\begin{tabular}{l|rrr|cc} \toprule
Dataset    			&\parbox{0.13\linewidth}{\centering\#users ($|\mathcal{U}|$)}    &\parbox{0.13\linewidth}{\centering\#items ($|\mathcal{I}|$)}  &\#actions   &\parbox{0.15\linewidth}{\centering avg. \#actions /user}  &\parbox{0.15\linewidth}{\centering avg. \#actions /item}  \\ \midrule 
\emph{Epinions}    &5,015       &8,335     &26,932      &5.37   &3.23  \\ 
\emph{Automotive}  &34,316      &40,287    &183,573     &5.35   &4.56  \\  
\emph{Google}      &350,811     &505,516   &2,591,026   &7.39   &5.13 \\
\emph{Office}      &16,716      &22,357    &128,070     &7.66   &5.73  \\ 
\emph{Toys}        &57,617      &69,147    &410,920     &7.13   &5.94  \\
\emph{Clothing}    &184,050     &174,484   &1,068,972   &5.81   &6.13  \\
\emph{Cellphone}   &68,330      &60,083    &429,231     &6.28   &7.14  \\
\emph{Games}       &31,013      &23,715    &287,107     &9.26   &12.11 \\
\emph{Electronics} &253,996     &145,199   &2,109,879   &8.31   &14.53 \\
\emph{Foursquare}  &43,110      &13,335    &306,553     &7.11   &22.99 \\ 
\emph{Flixter}     &69,485      &25,759    &8,000,971   &115.15 &310.61 \\ \hline
\textbf{Total}     &\textbf{1.11M} &\textbf{1.09M } &\textbf{15.5M} &{-}  &{-}   \\ \bottomrule
\end{tabular}
\end{center}
\end{table}

\xhdr{Amazon.}\footnote{\url{http://jmcauley.ucsd.edu/data/amazon/}} The first group of datasets, comprising large corpora of reviews and timestamps on various products, were recently introduced by \cite{VisualSIGIR}. These data are originally from \emph{Amazon.com} and span May 1996 to July 2014. Top-level product categories on \emph{Amazon} were constructed as separate datasets by \cite{VisualSIGIR}. In this paper, we take a series of large categories including `Automotive,' `Cell Phones and Accessories,' `Clothing, Shoes, and Jewelry,' `Electronics,' `Office Products,' `Toys and Games,' and `Video Games.' 
This set of data is notable for its high sparsity and variability.

\xhdr{Epinions.\footnote{\url{http://jmcauley.ucsd.edu/data/epinions/}}} This dataset was collected by \cite{zhao2014leveraging} from \emph{Epinions.com}, a popular online consumer review website. The reviews span January 2001 to November 2013.

\xhdr{Foursquare.\footnote{\url{https://archive.org/details/201309_foursquare_dataset_umn}}} Is originally from \emph{Foursquare.com}, containing a large number of check-ins of users at different venues from December 2011 to April 2012. This dataset was collected by \cite{levandoski2012lars} and is widely used for evaluating next point-of-interest prediction methods.

\xhdr{Flixter.\footnote{\url{http://www.cs.ubc.ca/~jamalim/datasets/}}} A large, dense movie rating dataset from \emph{Flixter.com}. The timespan is from November 2005 to November 2009.

\begin{figure}
\centering
\includegraphics[width=\linewidth]{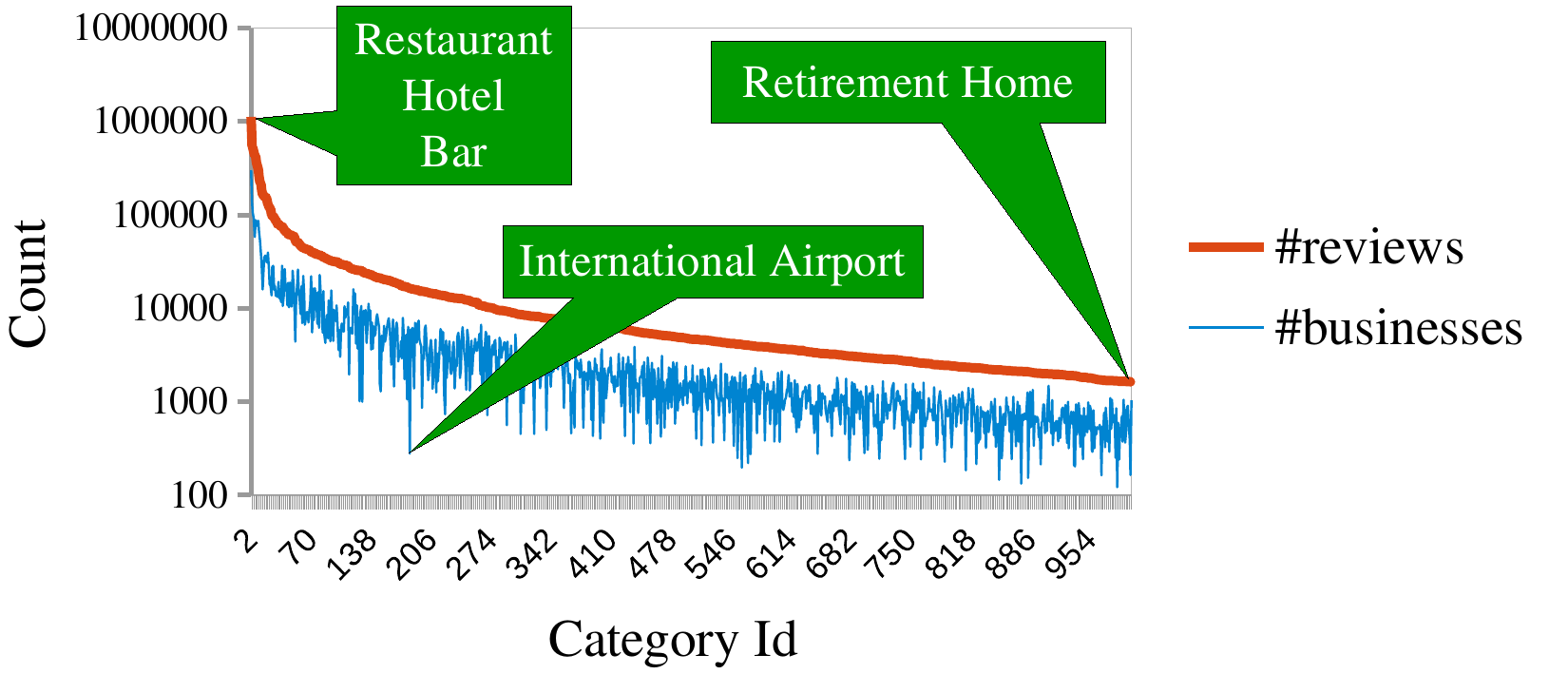}
\caption{Number of reviews and local businesses associated with the top 1,000 popular categories from \emph{Google Local}.}
\label{fig:google}
\end{figure}

\xhdr{Google Local.} We introduce a new dataset from \emph{Google} which contains 11,453,845 reviews and ratings from 4,567,431 users on 3,116,785 local businesses (with detailed name, hours, phone number, address, GPS, etc.). There are as many as 48,013 categories of local businesses distributed over five continents, ranging from restaurants, hotels, parks, shopping malls, movie theaters, schools, military recruiting offices, bird control, mediation services (etc.).
Figure \ref{fig:google} shows the number of reviews and businesses associated with each of the top 1,000 popular categories. The vast vocabulary of items, variability, and data sparsity make it a challenging dataset to examine the effectiveness of our model.
Although not the goal of our study, this is also a potentially useful dataset for location-based recommendation.


For each of the above datasets, we discard users and items with fewer than 5 associated actions in the system. In cases where star-ratings are available, we take all of them as users' positive feedback, since we are dealing with implicit feedback settings and care about purchases/check-in actions (etc.)~rather than the specific ratings. Statistics of our datasets (after pre-processing) are shown in Table \ref{tb:data}.

\subsection{Comparison Methods}
\xhdr{PopRec:} This is a na\"{\i}ve baseline that ranks items according to their popularity, i.e.,~it recommends the most popular items to users and is not personalized.

\xhdr{Bayesian Personalized Ranking (BPR-MF) \cite{rendle2009bpr}:} BPR-MF is a state-of-the-art item recommendation model which takes Matrix Factorization as the underlying predictor. It ignores the sequential signals in the system.

\xhdr{Factorized Markov Chain (FMC):} Captures the `global' sequential dynamics by factorizing the item-to-item transition matrix (shared by all users), but does not capture personalized behavior.

\xhdr{Factorized Personalized Markov Chain (FPMC) \cite{rendle2010fpmc}:} Uses a predictor that combines Matrix Factorization and factorized Markov Chains so that personalized Markov behavior can be captured (see \eq{eq:fpmc}).

\xhdr{Personalized Ranking Metric Embedding (PRME) \cite{feng2015prme}:} PRME models personalized Markov behavior by the summation of two Euclidean distances (see \eq{eq:prme}).

\xhdr{Hierarchical Representation Model (HRM) \cite{hrm}:} HRM extends FPMC by using aggregation operations like max pooling to model more complex interactions (see \eq{eq:hrm}). We compare against HRM with both max pooling and average pooling, denoted by HRM$_{\text{max}}$ and HRM$_{\text{avg}}$ respectively.

\xhdr{Translation-based Recommendation (\md):} Our method, which unifies user preferences and sequential dynamics with translations. In experiments we try both $\mathcal{L}_1$ and squared $\mathcal{L}_2$ distance
\footnote{Note that this can be seen as optimizing an $\mathcal{L}_2$ distance space, similar to the approach used by PRME \cite{feng2015prme}.}
for our predictor (see \eq{eq:TransRec}). 

Table \ref{tb:property} examines the properties of different methods. The ultimate goal of the baselines is to demonstrate (1) the performance achieved by state-of-the-art sequentially-unaware item recommendation models (BPR-MF) and purely sequential models without modeling {personalization} (FMC); (2) the benefits of combining personalization and sequential dynamics in a `linear' (FPMC) and non-linear way (HRM), or using metric embeddings (PRME); and (3) the strength of \md{} using {translations}.

\subsection{Evaluation Methodology} \label{sec:metric}
For each dataset, we partition the historical sequence $\mathcal{S}^u$ for each user $u$ into three parts: (1) the most recent one $\mathcal{S}^u_{|\mathcal{S}^u|}$ for test, (2) the second most recent one $\mathcal{S}^u_{|\mathcal{S}^u|-1}$ for validation, and (3) all the rest for training. Hyperparameters in all cases are tuned by grid search with the validation set. Finally, we report the performance of each method on the test set in terms of the following ranking metrics:

\xhdr{Area Under the ROC Curve (AUC):}
$$
\mathit{AUC} =  \frac{1}{|\mathcal{U}|}  \sum_{u \in \mathcal{U}}   \frac{1}{|\mathcal{I}\setminus \mathcal{S}^u|}   \sum_{j' \in \mathcal{I} \setminus \mathcal{S}^u}  \mathbf{1} (R_{u,g_u} < R_{u,j'}),
$$

\xhdr{Hit Rate at position 50 (Hit@50):}
$$
\mathit{Hit@50} =  \frac{1}{|\mathcal{U}|}  \sum_{u \in \mathcal{U}}  \mathbf{1}(R_{u,g_u} \leq 50),
$$
where $g_u$ is the `ground-truth' item associated with user $u$ at the most recent time step,
$R_{u,i}$ is the rank of item $i$ for user $u$ (smaller is better), and $\mathbf{1}(b)$ is an indicator function that returns $1$ if the argument $b$ is $\mathit{true}$; $0$ otherwise.

\begin{table}
\begin{center}
\setlength{\tabcolsep}{1.6pt}
\caption{Models. {P: Personalized? S: Sequentially-aware? M: Metric-based?
U: Unified model of third-order relations?}}\label{tb:property}
\begin{tabular}{cccccccccc} \toprule
Property    &PopRec      &BPR-MF            &FMC            &FPMC           &HRM             &PRME            &\md{}            \\ \midrule 
{P}  &\ding{56}    &\ding{52}      &\ding{56}   &\ding{52} &\ding{52}  &\ding{52}  &\ding{52} \\  
{S}  &\ding{56}    &\ding{56}      &\ding{52} &\ding{52} &\ding{52}  &\ding{52}  &\ding{52} \\
{M}  &\ding{56}    &\ding{56}      &\ding{56}   &\ding{56}   &\ding{56}    &\ding{52}  &\ding{52} \\
{U}  &\ding{56}    &\ding{56}      &\ding{56}   &\ding{56}   &\ding{52}  &\ding{56}    &\ding{52} \\  \bottomrule
\end{tabular}
\end{center}
\end{table}

\begin{table*}
\centering
\setlength{\tabcolsep}{2pt}
\caption{Ranking results on different datasets (higher is better). The number of latent dimensions $K$ for all comparison methods is set to 10. The best performance in each case is underlined. The last column shows the percentage improvement of \md{} over the best baseline. 
}
\begin{tabular}{l|c|cccccccccr} \toprule
Dataset  &\parbox{1.2cm}{\centering Metric}      &\parbox{1.3cm}{\centering PopRec}  &\parbox{1.3cm}{\centering BPR-MF} &\parbox{1.3cm}{\centering FMC} &\parbox{1.3cm}{\centering FPMC} &\parbox{1.3cm}{\centering HRM$_{\text{avg}}$}   &\parbox{1.3cm}{\centering HRM$_{\text{max}}$} &\parbox{1.3cm}{\centering PRME} &\parbox{1.4cm}{\centering TransRec$_{\mathcal{L}_1}$}    &\parbox{1.5cm}{\centering TransRec$_{\mathcal{L}_2}$} &\parbox{1.3cm}{\centering \%Improv.}  \\ \midrule
\multirow{2}{*}{\emph{Epinions}}    &\emph{AUC} 	 	&0.4576  &0.5523  &0.5537  &0.5517  &0.6060  &0.5617  &0.6117  &0.6063  &\bul{0.6133}   &0.3\%     \\
							        &\emph{Hit@50}	&3.42\%  &3.70\%  &3.84\%  &2.93\%  &3.44\%  &2.79\%  &2.51\%  &3.18\%  &\bul{4.63\%}   &20.6\%     \\ [1.5pt]
                               
\multirow{2}{*}{\emph{Automotive}}  &\emph{AUC} 	 	&0.5870  &0.6342  &0.6438  &0.6427  &0.6704  &0.6556  &0.6469  &0.6779  &\bul{0.6868}   &2.5\%     \\
							        &\emph{Hit@50}	&3.84\%  &3.80\%  &2.32\%  &3.11\%  &4.47\%  &3.71\%  &3.42\%  &5.07\%  &\bul{5.37\%}   &20.1\%     \\ [1.5pt]

\multirow{2}{*}{\emph{Google}}      &\emph{AUC} 	 	&0.5391  &0.8188  &0.7619  &0.7740  &0.8640  &0.8102  &0.8252  &0.8359  &\bul{0.8691}   &0.6\%      \\
							        &\emph{Hit@50}	&0.32\%  &4.27\%  &3.54\%  &3.99\%  &3.55\%  &4.59\%  &5.07\%  &6.37\%  &\bul{6.84\%}   &32.5\%     \\ [1.5pt]

\multirow{2}{*}{\emph{Office}}      &\emph{AUC} 	 	&0.6427  &0.6979  &0.6867  &0.6866  &0.6981  &0.7005  &0.7020  &0.7186  &\bul{0.7302}   &4.0\%     \\
							        &\emph{Hit@50}	&1.66\%  &4.09\%  &2.66\%  &2.97\%  &5.50\%  &4.17\%  &6.20\%  &\bul{6.86\%}  &6.51\%   &10.7\%     \\ [1.5pt]

\multirow{2}{*}{\emph{Toys}}        &\emph{AUC} 	 	&0.6240  &0.7232  &0.6645  &0.7194  &0.7579  &0.7258  &0.7261  &0.7442  &\bul{0.7590}   &0.2\%     \\
							        &\emph{Hit@50}	&1.69\%  &3.60\%  &1.55\%  &4.41\%  &5.25\%  &3.74\%  &4.80\%  &\bul{5.46\%}  &5.44\%   &4.0\%     \\ [1.5pt]
                              
\multirow{2}{*}{\emph{Clothing}}    &\emph{AUC} 	 	&0.6189  &0.6508  &0.6640  &0.6646  &0.7057  &0.6862  &0.6886  &0.7047  &\bul{0.7243}   &2.6\%     \\
							        &\emph{Hit@50}	&1.11\%  &1.05\%  &0.57\%  &0.51\%  &1.70\%  &1.15\%  &1.00\%  &1.76\%  &\bul{2.12\%}   &24.7\%     \\ [1.5pt]

\multirow{2}{*}{\emph{Cellphone}}   &\emph{AUC} 	 	&0.6959  &0.7569  &0.7347  &0.7375  &0.7892  &0.7654  &0.7860  &0.7988  &\bul{0.8104}   &2.7\%     \\
							        &\emph{Hit@50}	&4.43\%  &5.15\%  &3.23\%  &2.81\%  &8.77\%  &6.32\%  &6.95\%  &9.46\%  &\bul{9.54\%}   &8.8\%     \\ [1.5pt]
                       
\multirow{2}{*}{\emph{Games}}       &\emph{AUC} 	 	&0.7495  &0.8517  &0.8407  &0.8523  &0.8776  &0.8566  &0.8597  &0.8711  &\bul{0.8815}   &0.4\%     \\
							        &\emph{Hit@50}	&5.17\%  &10.93\% &13.93\% &12.29\% &14.44\% &12.86\% &14.22\% &\bul{16.61\%} &16.44\%  &15.0\%     \\ [1.5pt]
                               
\multirow{2}{*}{\emph{Electronics}} &\emph{AUC} 	 	&0.7837  &0.8096  &0.8158  &0.8082  &0.8212  &0.8148  &0.8337  &0.8457  &\bul{0.8484}   &1.8\%     \\
							        &\emph{Hit@50}	&4.62\%  &2.98\%  &4.15\%  &2.82\%  &4.09\%  &2.59\%  &3.07\%  &4.89\%  &\bul{5.19\%}   &12.3\%     \\ [1.5pt]
                               
\multirow{2}{*}{\emph{Foursquare}}  &\emph{AUC} 	 	&0.9168  &0.9511  &0.9463  &0.9479  &0.9559  &0.9523  &0.9565  &0.9631  &\bul{0.9651}   &0.9\%     \\
							        &\emph{Hit@50}	&55.60\% &60.03\% &63.00\% &64.53\% &60.75\% &61.60\% &65.32\% &66.12\% &\bul{67.09\%}  &2.7\%     \\ [1.5pt]                             
                              
\multirow{2}{*}{\emph{Flixter}}     &\emph{AUC} 	 	&0.9459  &0.9722  &0.9568  &0.9718  &0.9695  &0.9687  &0.9728  &0.9727  &\bul{0.9750}   &0.2\%      \\
							        &\emph{Hit@50}	&11.92\% &21.58\% &22.23\% &33.11\% &32.34\% &30.88\% &\bul{40.81\%} &35.52\% &35.02\%  &-13.0\%    \\ \bottomrule
\end{tabular}
\label{tb:i2u}
\end{table*}

\subsection{Performance and Quantitative Analysis}
Results are collated in Table \ref{tb:i2u}. Due to the sparsity of most of the datasets in consideration, the number of dimensions $K$ of all latent vectors in all cases is set to 10 for simplicity; 
we investigate the importance of the number of dimensions in our parameter study later.
Note that in Table \ref{tb:i2u} datasets are ranked in ascending order of item density. The last column (\%Improv.) demonstrates the percentage improvement of \md{} over the strongest baseline for each dataset. The main findings are summarized as follows:

BPR-MF and FMC achieve considerably better results than the popularity-based baseline in most cases, in spite of modeling personalization and sequential patterns in isolation. This means that uncovering the underlying user-item and item-item relationships is key to making meaningful recommendations. 

FPMC and HRM are essentially combinations of MF and FMC. FPMC beats BPR-MF and FMC mainly on relatively dense datasets like \emph{Toys}, \emph{Foursquare}, and \emph{Flixter}, and loses on sparse datasets---possibly due to the large number of parameters it introduces. From Table \ref{tb:i2u} we see that HRM achieves strong results amongst all baselines in most cases, presumably from the aggregation operations.

PRME replaces the inner products in FPMC by distance functions. It beats FPMC in most cases, though loses to HRM due to different modeling strategies. Note that like FPMC, PRME turns out to be quite strong at handling dense datasets like \emph{Foursquare} and \emph{Flixter}. We speculate that the two models could benefit from the considerable amount of additional parameters they use when data is dense.

\md{} outperforms other methods in nearly all cases. The improvements seem to be correlated with:

\textbf{Variability.} \md{} achieves large improvements (32.5\% and 24.7\% in terms of {Hit@50}) on \emph{Google} and \emph{Clothing}, two datasets with the largest vocabularies of items in our collection. Taking \emph{Google} as an example, it includes all kinds of restaurants, bars, shops (etc.) as well as a global user base, which requires the ability to handle the vast variability. 

\textbf{Sparsity.} \md{} beats all baselines especially on comparatively sparser datasets like \emph{Epinions}, \emph{Automotive}, and \emph{Google}. The only exception is in terms of {Hit@50} on \emph{Flixter}, the \emph{densest} dataset in consideration. We speculate that \md{} is at a disadvantage by using fewer parameters (than PRME) especially when $K$ is set to a small number (10). As we demonstrate in Section \ref{sec:sensitivity}, we can achieve comparable results with the strongest baseline when increasing the model dimensionality.

In addition, we empirically find that (squared) $\mathcal{L}_2$ distance
typically outperforms $\mathcal{L}_1$ distance, though the latter also beats baselines in most cases.

\begin{figure*}
\centering
\includegraphics[width=\linewidth]{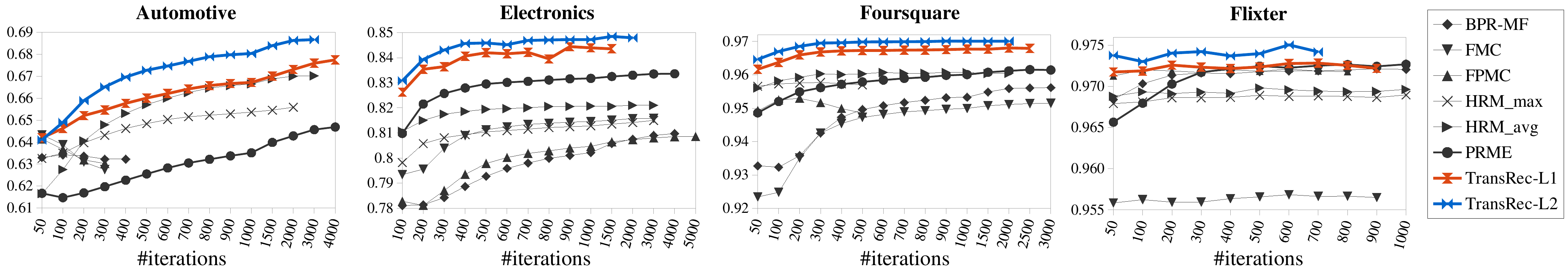}
\caption{Convergence: Test AUCs on four datasets as the number of training iterations increases ($K=10$).}
\label{fig:iter}
\end{figure*}

\begin{figure*}
\centering
\includegraphics[width=\linewidth]{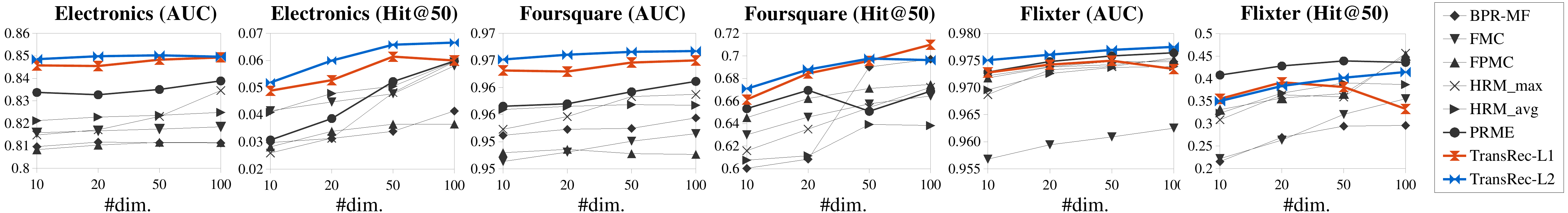}
\caption{Sensitivity: Accuracy variation on the three densest datasets with increasing dimensionality (i.e.,~$K$).}
\label{fig:dim}
\end{figure*}

\subsection{Convergence}
In Figure \ref{fig:iter} we demonstrate (test) AUCs with increasing training iterations on four datasets with varying sparsity---\emph{Automotive}, \emph{Electronics}, \emph{Foursquare}, and \emph{Flixter}. \emph{Automotive} is representative of sparse datasets in our collection. Simple baselines like FMC and BPR-MF converge faster than other methods on sparse datasets, presumably due to the relatively simpler dynamics they capture. FPMC also converges fast on such datasets as a result of its tendency to overfit 
(recall that we terminate once no further improvements are achieved on the validation set).
On denser datasets like \emph{Electronics}, \emph{Foursquare}, and \emph{Flixter}, all methods tend to converge at comparable speeds due to the need to unravel denser relationships amongst different entities.

\subsection{Sensitivity} \label{sec:sensitivity}
For the three densest datasets---\emph{Electronics}, \emph{Foursquare}, and \emph{Flixter}---we also experimented with different numbers of dimensions for user/item representations. We increase $K$ from 10 to 100 and present AUC and Hit@50 values on the test set in Figure \ref{fig:dim}.
\md{} still dominates other methods on \emph{Electronics} and \emph{Foursquare}. As for \emph{Flixter}, from the rightmost subfigure we can see that in terms of Hit@50 the gap between \md{} ($\mathcal{L}_2$) and PRME, the strongest baseline on this data, closes as we increase the dimensionality. 

\subsection{Implementation Details}
To make fair comparisons, we used stochastic gradient ascent to optimize pairwise rankings for all models (except PopRec) with a fixed learning rate of 0.05. Regularization hyperparamters are selected from $\{0, 0.001, 0.01, 0.1, 1\}$ (using the validation set).
We did not make use of the dropout technique mentioned in the HRM paper to make it comparable to other methods. For PRME, we selected $\alpha$ from $\{0.2, 0.5, 0.8\}$. 0.2 was found to be the best in the PRME paper, which is consistent with our own observations. For \md, we used the unit $\mathcal{L}_2$-ball as our subspace $\Psi$. We also tried using the unit $\mathcal{L}_2$-sphere
(i.e.,~the surface of the ball), but it led to slightly worse results in practice.

\subsection{Recommendations}
\begin{figure}
\centering
\includegraphics[width=\linewidth]{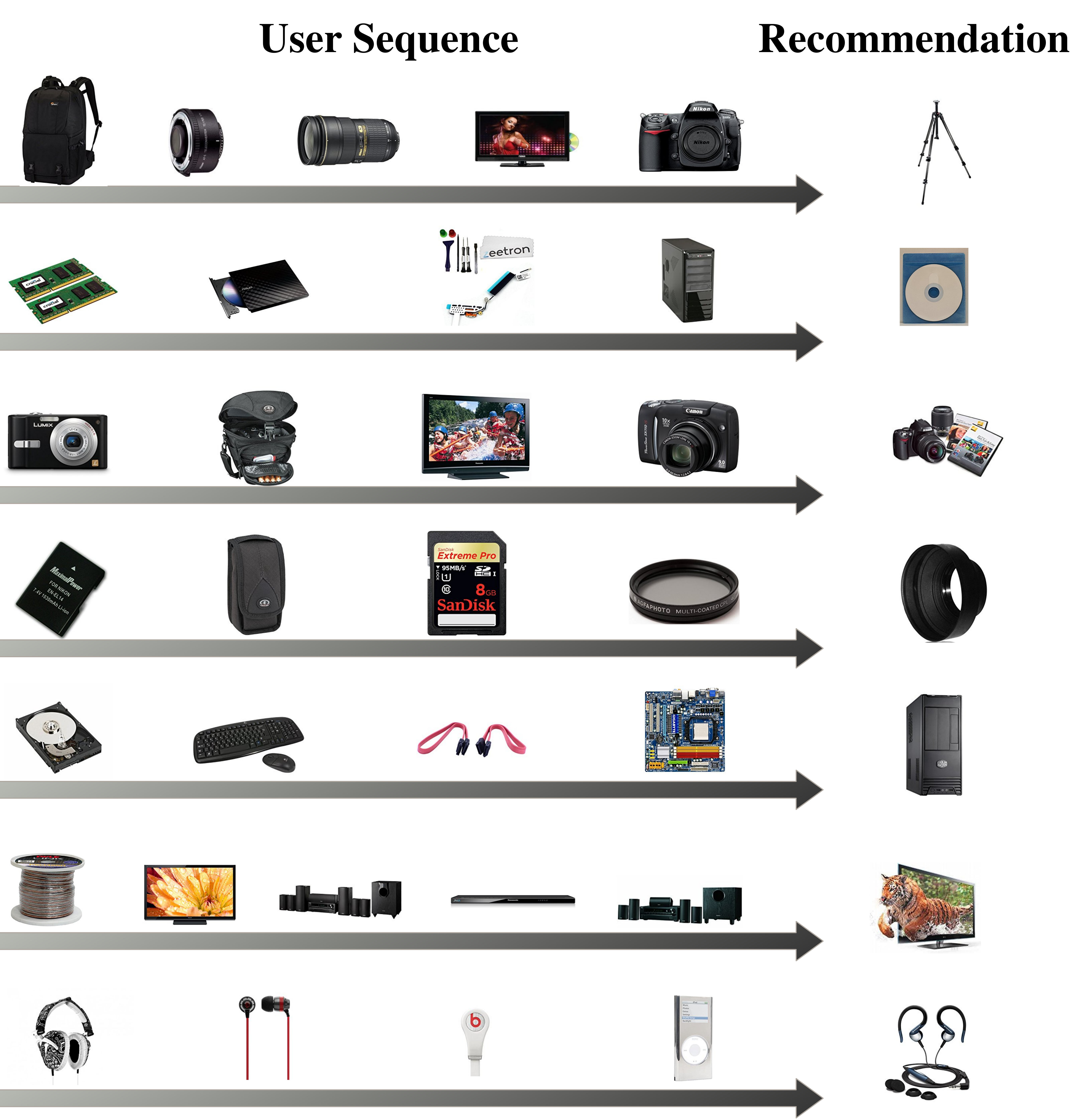}
\caption{Recommendations made for a random sample of seven users by \md{} on \emph{Electronics} data.}
\label{fig:rec}
\end{figure}

In Figure \ref{fig:rec} we demonstrate some recommendations made by \md{} ($K=10$) on Electronics. We randomly sample a few users from the datasets and show their historical sequences on the left, and demonstrate the top-1 recommendation on the right. As we can see from these examples, \md{} can capture long-term dynamics successfully. For example, \md{} recommends a tripod to the first user who appears to be a photographer. The last user bought multiple headphones and similar items in history;
\md{} recommends new headphones after the purchase of an iPod accessory. 
In addition, \md{} also captures short-term dynamics. For instance, it recommends a desktop case to the fifth user after 
the purchase of a motherboard. Similarly, the sixth user is recommended a HDTV after recently purchasing a home theater receiver/speaker.

\subsection{Item-to-item recommendation}
By removing the personalization element, \md{} can straightforwardly be adapted to handle item-to-item recommendation,
another classical setting where recommendations are made in the context of a specific item, e.g.,~recommending items that are likely to be purchased together. This setting is analogous to the knowledge graph completion task in that relationships among different items need to be modeled.

\subsubsection{Datasets and Evaluation Methodology}
  
 We use 8 large datasets representing \emph{co-purchase} relationships between products from \emph{Amazon} \cite{VisualSIGIR}. They are a variety of top-level \emph{Amazon} categories; to make the task more challenging, we only consider edges that connect two different top-level subcategories within each of the above datasets (e.g.,~recommending complementary items rather than substitutes). Statistics of these datasets are collated in Table \ref{tb:i2i-data}.

Note that in these datasets edges are \emph{directed}, e.g.,~it makes sense to recommend a charger/backpack after a customer purchases a laptop, but not the other way around.

\xhdr{Features.} To further evaluate \md{}, we consider testing its capability here as a \emph{content}-based method. To this end, 
we extract Bag-of-Words (BoW) features from each product's review text. In short, for each dataset we removed stop-words and constructed a dictionary comprising the 5,000 most frequent nouns or adjectives or adjective-noun bigrams. These features have been shown to be effective on this data \cite{he2016monomer}.

\begin{table}
\centering
\setlength{\tabcolsep}{4.3pt}
\caption{Statistics (in ascending order of \#edges). }\label{tb:i2i-data}
\begin{tabular}{llrrcc} \toprule
Dataset            &Full name          &\#items     &\#edges    \\ \midrule 
\emph{Office}      &Office Products    &130,006     &52,942     \\ 
\emph{Home}        &Home \& Kitchen    &410,244     &122,955    \\  
\emph{Games}       &Video Games        &50,210      &314,124    \\
\emph{Electronics} &Electronics        &476,004     &549,914    \\ 
\emph{Automotive}  &Automotive         &320,116     &637,814    \\
\emph{Movies \& TV}&Movies \& TV       &200,941     &648,256    \\
\emph{Cellphone}   &Cell Phones \& Accessories &319,678     &667,918    \\
\emph{Toys}        &Toys \& Games      &327,699      &948,729    \\ \hline
\textbf{Total}     &                   &\textbf{2.23M} &\textbf{3.94M }    \\ \bottomrule
\end{tabular}
\end{table}

\xhdr{Evaluation Methodology.} For each of the above datasets, we randomly partition the edges with an 80\%/10\%/10\% train/validation/test split. Validation is used to select hyperparameters and performance is reported on the test set. Again we report AUC and Hit@10 (see Section \ref{sec:metric}). Here we use 10 for the hit rate because, as we show later, item-to-item recommendation proves simpler than personalized sequential prediction.

\subsubsection{The Translation-based Model}
Here we adopt a \emph{content}-based version of \md, to investigate its ability to tackle explicit features. Let $\vec{f}_i$ denote the explicit feature vector associated with item $i$. We add one additional embedding layer $E(\cdot)$ on top of $\vec{f}$ to project items into the `relational space' $\Phi$. Formally, \md{} makes predictions according to
$$
\begin{aligned}
\mathit{Prob}(j~|~i) \propto  -\; d \left( E(\vec{f}_i) + \vec{t}, E(\vec{f}_j) \right), \\
\text{subject to} \quad E(\vec{f}_i) \in \Psi \subseteq \Phi, \;\;\; \text{for}{\ } i \in \mathcal{I}.
\end{aligned}
$$
$E(\cdot)$ could be a linear embedding layer, a non-linear layer like a neural network, or even some combination of latent and content-based representations.

\subsubsection{Baselines}
We mainly compare against two related models based on metric (or \emph{non}-metric) embeddings. These are state-of-the-art \emph{content}-based methods for item-to-item recommendation and have demonstrated strong results on the same data \cite{VisualSIGIR,he2016monomer}. The complete list of baselines is as follows:

\xhdr{Weighted Nearest Neighbor (WNN):} WNN measures the `dissimilarity' between pairs of items by a weighted Euclidean distance in the raw feature space: 
$d_{\vec{w}}(i, j) = \| \vec{w} \circ (\vec{f}_i - \vec{f}_j) \|_2^2$, where $\circ$ is the Hadamard product and $\vec{w}$ is a parameter to be learned. 

\xhdr{Low-rank Mahalanobis Transform (LMT) \cite{VisualSIGIR}:} A state-of-the-art embedding method for learning the notion of compatibilities among different items. LMT learns a single low-rank Mahalanobis transform
matrix $W$ to embed all items into a relational space within which the distance between items is measured to make predictions: $d_W(i, j) = \| W \vec{f}_i - W \vec{f}_j \|_2^2$.  

\xhdr{Mixtures of Non-metric Embeddings (Monomer) \cite{he2016monomer}:} Monomer extends LMT by learning \emph{mixtures} of low-rank embeddings to uncover more complex reasons to explain the relationships between items. It relaxes the metricity assumption used by LMT and can naturally handle directed relationships.

\subsubsection{Quantitative Results and Analyses}
For fair comparison, we adopted the setting in \cite{he2016monomer}, 
so that
we use 100 dimensions for the relational spaces of LMT and \md; 5 spaces each with 20 dimensions are learned for Monomer. For simplicity, in our experiments we used squared $\mathcal{L}_2$ distance and $\Psi = \Phi$ for \md{}, i.e.,~no 
constraints on the vector $E(\vec{f})$. Also, a linear embedding layer is used as the function $E$ to make it more comparable with our baselines. 

Experimental results are collated in Table \ref{tb:i2i}. Our main findings are summarized as follows: (1) \md{} outperforms all baselines in all cases considerably, which indicates that translation-based structure seems to be stronger at modeling relationships among items compared to purely distance-based methods. This is also consistent with the findings from knowledge base literature (e.g.,~\cite{TransE,TransH,TransR}). (2) \md{} tends to lead to larger improvements for sparse datasets like \emph{Office}, in contrast to the improvements on denser datasets like \emph{Toys} and \emph{Games}. 

\begin{table}
\centering
\setlength{\tabcolsep}{2.4pt}
\caption{Accuracy for co-purchase prediction (higher is better).}
{\small
\begin{tabular}{l|c|ccccr} \toprule
Dataset &Metric &WNN &LMT  &Monomer  &TransRec  &\%Improv.  \\ \midrule

\multirow{2}{*}{\emph{Office}}      &\emph{AUC}    &0.6952  &0.8848  &0.8736  &0.9437   &6.7\%    \\
							        &\emph{Hit@10} &1.45\%  &3.08\%  &1.96\%  &12.69\%  &312.0\%  \\ [1.5pt]
\multirow{2}{*}{\emph{Home}}        &\emph{AUC}    &0.6696  &0.9101  &0.8841  &0.9482   &4.2\%    \\
									&\emph{Hit@10} &2.24\%  &4.46\%  &0.63\%  &8.80\%   &97.3\%   \\ [1.5pt]
\multirow{2}{*}{\emph{Games}}       &\emph{AUC}    &0.7199  &0.9423  &0.9239  &0.9736   &3.3\%    \\
									&\emph{Hit@10} &2.64\%  &4.19\%  &0.59\%  &7.78\%   &85.7\%   \\ [1.5pt] 
\multirow{2}{*}{\emph{Electronics}} &\emph{AUC}    &0.7538  &0.9316  &0.9299  &0.9651   &3.5\%    \\
									&\emph{Hit@10} &1.78\%  &2.59\%  &0.29\%  &5.32\%   &105.4\%  \\ [1.5pt]
\multirow{2}{*}{\emph{Automotive}}  &\emph{AUC}    &0.7317  &0.9054  &0.9152  &0.9490   &3.7\%    \\
									&\emph{Hit@10} &1.20\%  &1.97\%  &0.36\%  &4.48\%   &127.4\%  \\ [1.5pt]
\multirow{2}{*}{\emph{Movies \& TV}}&\emph{AUC}    &0.7668  &0.9536  &0.9516  &0.9730   &1.9\%    \\
									&\emph{Hit@10} &2.84\%  &4.37\%  &0.99\%  &6.19\%   &41.7\%   \\ [1.5pt]
\multirow{2}{*}{\emph{Cellphone}}   &\emph{AUC}    &0.6867  &0.7932  &0.8445  &0.9127   &8.1\%    \\
									&\emph{Hit@10} &0.80\%  &0.94\%  &0.04\%  &2.42\%   &157.5\%  \\ [1.5pt]
\multirow{2}{*}{\emph{Toys}}        &\emph{AUC}    &0.7529  &0.9216  &0.9353  &0.9552   &2.1\%    \\
									&\emph{Hit@10} &2.27\%  &2.67\%  &0.59\%  &3.99\%   &49.4\%   \\ 	\bottomrule
\end{tabular}
}\normalsize
\label{tb:i2i}
\end{table} 

\section{Conclusion}
We introduced a scalable \emph{translation}-based method, \md, for modeling the semantically complex relationships between different entities in recommender systems. We analyzed the connections of \md{} to existing methods and demonstrated its suitability for modeling third-order interactions between users, their previously consumed item, and their next item. 
In addition to the superior results achieved on the sequential prediction task on a wide spectrum of large, real-world datasets, we also investigated the strength of \md{} at tackling item-to-item recommendation. 
The success of \md{} on the two tasks suggests that translation-based architectures are promising for general-purpose recommendation problems.

In addition, we introduced a large-scale dataset for sequential (and potentially geographical) recommendation from \emph{Google Local}, that contains detailed information about millions of local businesses (e.g.,~restaurants, malls, shops) around the world as well as ratings and reviews from millions of users.

\bibliographystyle{ACM-Reference-Format}
\bibliography{sigproc} 
\end{document}